\documentclass[11pt]{article}
\pdfoutput=1
\usepackage[utf8]{inputenc}

\usepackage{graphicx}
\usepackage{subfig}
\usepackage{amsmath,amssymb}
\usepackage{float}
\usepackage{rotating}
\usepackage{geometry}
\usepackage[round]{natbib}
\usepackage{hyperref}
\usepackage{xcolor}
\definecolor{dark-red}{rgb}{0.4,0.15,0.15}
\definecolor{dark-blue}{rgb}{0.15,0.15,0.8}
\definecolor{medium-blue}{rgb}{0,0,0.5}
\hypersetup{
    colorlinks, linkcolor={dark-red},
    citecolor={dark-blue}, urlcolor={medium-blue}
}
\usepackage{multirow}
\usepackage{graphpap}

\begin{document}

\title{\textit{flora robotica} -- An Architectural System\\ Combining Living Natural Plants and Distributed Robots}

\author{
Heiko Hamann,\thanks{Corresponding author, e-mail: {\tt\small hamann@iti.uni-luebeck.de}}\hspace{.5mm}~\footnote{Institute of Computer Engineering, University of L\"ubeck, Germany}
~Mohammad Divband Soorati,$^\dagger$ 
Mary Katherine Heinrich,\footnote{Centre for IT and Architecture, Royal Danish Academy (KADK), Copenhagen, Denmark} \\
~Daniel Nicolas Hofstadler,\footnote{Artificial Life Lab of the Department of Zoology, Karl-Franzens University Graz, 
Austria}
~Igor Kuksin,\footnote{Cybertronica UG, Stuttgart, Germany}
~Frank Veenstra,\footnote{Robotics, Evolution and Art Lab (REAL), IT University of Copenhagen,
Denmark} \\
~Mostafa Wahby,$^\dagger$ 
Stig Anton Nielsen,$^\|$ 
Sebastian Risi,$^\|$ \\
Tomasz Skrzypczak,\footnote{Dept. of Molecular and Cellular Biology, Adam Mickiewicz University, Pozna{\'n}, Poland} 
~~Payam Zahadat,$^\mathsection$ 
Przemyslaw Wojtaszek,$^{**}$ \\
Kasper St{\o}y,$^\|$ 
Thomas Schmickl,$^\mathsection$ 
Serge Kernbach,$^\mathparagraph$ 
Phil Ayres$^\ddagger$
}
\date{}
\maketitle

\begin{abstract}
  Key to our project \textit{flora robotica} is the idea of creating a bio-hybrid system of tightly coupled natural plants and distributed robots to grow architectural artifacts and spaces. Our motivation with this ground research project is to lay a principled foundation towards the design and implementation of living architectural systems that provide functionalities beyond those of orthodox building practice, such as self-repair, material accumulation and self-organization. Plants and robots work together to create a living organism that is inhabited by human beings. User-defined design objectives help to steer the directional growth of the plants, but also the system's interactions with its inhabitants determine locations where growth is prohibited or desired (e.g., partitions, windows, occupiable space). We report our plant species selection process and aspects of living architecture. A~leitmotif of our project is the rich concept of braiding: braids are produced by robots from continuous material and serve as both scaffolds and initial architectural artifacts before plants take over and grow the desired architecture. We use light and hormones as attraction stimuli and far-red light as repelling stimulus to influence the plants. Applied sensors range from simple proximity sensing to detect the presence of plants to sophisticated sensing technology, such as electrophysiology and measurements of sap flow. We conclude by discussing our anticipated final demonstrator that integrates key features of \textit{flora robotica}, such as the continuous growth process of architectural artifacts and self-repair of living architecture. 

\end{abstract}

\newpage

\section{Introduction}

Growing living architecture in a bio-hybrid system based on natural plants and robots comes with a number of advantages compared to current housing construction techniques. The addition of material and the material itself is cheap, logistics are minor as the material grows at the site. Hence, energy expenditure for transport or production of building material is greatly lessened. The construction (growth) process is continuous, allowing the system to indefinitely compensate for wear or changed requirements, such as heavier loads on a bridge. The creation of pleasant and healthy environments for habitation is supported by natural components, such as trees and flowers, in lieu of concrete, bricks, or plaster~\citep{lee11}.
Fighting air and noise pollution but also an increasing heat~\citep{weber15,oliveira11} in our ever growing mega-cities is simplified once buildings by default consist of so-called `green infrastructure'~\citep[e.g., green walls, green roofs, ][]{sandstroem02,tzoulas2007promoting} that possibly absorbs pollutants and noise. A~living house can react to its inhabitants, was potentially evolved for resilience, and adapts to changed needs. Hence, living architecture based on plants~\citep{shankar2015living,gale2011potential,ludwig2016designing} can possibly contribute solutions to the fundamental global challenges of climate change and a growing (city) population.

At the same time, these ideas also come with scientific challenges and high demands to biology (in particular plant science), architecture, robotics, and engineering~\citep{hamann15b,hamann16a}. Appropriate plant species need to be selected according to their speed of growth, their growing type (climbing, bush, tree, annual/perennial, average leaf size, etc.), and the accumulated knowledge about them (model plants). Plants differ in their reaction to light, mechanical interaction, and hormones. Also mutants of plants could be considered, for example, if their growth turns out to be easier to steer.
The architectural challenges of a continuous growth process are much more complex, have many more degrees of freedom, need new types of design rules, and possibly even a new mindset opposing architecture that has a well defined endpoint in time. Plants in living architecture can be used directly as building material but also to change material properties of other, `artificial' building material. For example, plants can be grown through other materials to make them stiffer. During the initial explorative period of our project we have identified braids as a methodology with overwhelming opportunities. Braided structures can be produced autonomously by robots at site, can be used as scaffolds, can be stiffened by plants, but we can also embed electronics, sensors, actuators, computational units, and robots into strips and braids. The braids can solve the problem of slow plant growth~\citep{heinrich16}. Instead of having to wait for decades to move into the living house, the house can start from an autonomously constructed `dead' braided structure that is then successively substituted by grown alive plant material. 
We use generative and developmental methods to guide the growth and replacement pattern of these braided structures~\citep{zahadat17}. They combine environmental and internal sensory information and consider local and global requirements of the living architectural artifact.
A~braiding robot can serve as a construction robot only but it can also stay at the site as part of the living building to continue to produce braided scaffolds adaptively to the plants' needs and growth~\citep[similar to a self-assembly process, ][]{divband16a}.

The requirements for the robots that steer plants and interact with them are also rather complex. The robots need appropriate actuators to provide stimuli that trigger the natural adaptive behavior of plants. In \textit{flora robotica}, we use visible light to attract plants and far-red light to repel plant growth~\citep{florarobotica}. Using light may be challenging in outdoor conditions but fighting the sun may not be required when working with blue light possibly supported by devices to shade parts of the construction. Options of mobile robots seem to require climbing capabilities that we do not explore in \textit{flora robotica}. Instead, the robot design is simplified by keeping robots static while they can be replaced via user interaction. Actually, that is a suitable option given the slow speed of plant growth, hence, making replacements necessary only infrequently. This is in line with a long ignored alternative to standard robotics, namely the concept of `slow bots'~\citep{arkin15}. 

Sensing is an exceptional challenge in \textit{flora robotica}, starting from sensing the mere presence of plants to measuring the plant's well-being based on sap flow and electrophysiology. Following the idea of phytosensing~\citep{mazarei08}, we can use plants to report certain environmental factors, that may be difficult to sense with technological sensors. Then the challenge, in turn, is to sense the respective reaction of the plant. 

The control of plant-steering robots itself can be simple if only predetermined patterns need to be grown. However, the combined plant-robot experiments need to consolidate protocols and traditions from two very different fields. Even simple features, such as lab room temperature and light conditions, need to be selected wisely and need to compromise requirements of robots, plants, and human beings. The robot control gets complex once only certain percentages of desired biomass are determined by the user or once also the motion of plants needs to be controlled. In a recent paper we describe our methodology of machine learning techniques (evolutionary robotics) to steer plant tips into desired targets~\citep{wahby16a}.

The philosophy of bio-hybrid plant-robot living architecture has a lot of potential and could be pushed further to having equal roles of plants and robots resembling a symbiosis (e.g., also the plants are allowed to request replacements of robots or to request actuation). An interesting research track could be to develop growing robots that behave similarly to plants~\citep{mazzolai14,mazzolai10,hawkes17} and one could even try to control plants such that they complete tasks that were previously reserved for robots. This could include grasping or to navigate plants towards connection positions and connecting building material by growing around them~\citep[alive plants as connections, similar to alive ants for stitching, ][]{gudger25}. 

In \textit{flora robotica} we focus on developing a generic basic methodology for a radically new field of plant-robot living architecture. We plan to showcase our achievements in a benchmark where we grow a small wall with two holes cut into it, where one hole is supposed to be a window (growth prohibited) and one represents damage and is to be regrown. Hence, we would like to showcase a living architecture with the capability of self-repair and continuous growth.

\vspace{2mm}
\section{Natural plants}
\begin{figure}[t]
\includegraphics[width=\textwidth]{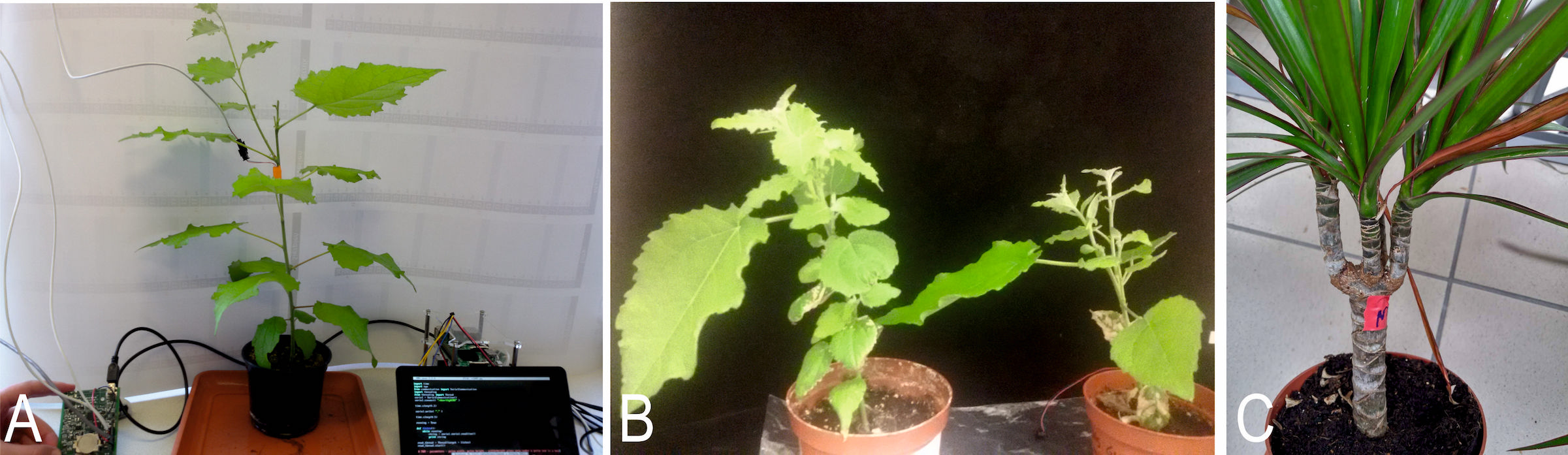}
\caption{\label{fig:amu}Examples of plant shaping methods: A) vibration motor 
that put the poplar into vibration for 5 minutes every hour; B) differences in rate of growth between a plant that was stimulated by vibration (right) and a control plant (left); C) stem of \textit{Dracaena marginata}, the marked branch was sprayed with NAA (auxin). The curvature of the branch has changed in 10 days.}
\end{figure}

Plants are suitable candidates to be explored in the domain of living architecture because they could be used as an additive manufacturing system similar to 3d printing. Besides growing mere structures, plants show adaptive behavior that could be used as a feature in living architecture. The life form of sessile plants comes with great abilities to adapt to locally determined conditions. Also the shape of a plant, as obtained by growth, shows developmental plasticity. A~plant's shape depends on an integration of past environmental conditions, the history of its life, and its genotype. All these factors allow the plant to make efficient use of local resources and to resist to local disturbances and stresses. Shapes of the root system and of the organs above ground is not only determined by environmental factors but is also restricted by pattern formation and topological preferences specific to the species. Obviously, we have accumulated a lot of knowledge to arrange and shape plants in the culture of gardening. However, instead of using this highly developed handcraft, that considers plants rather as a piece of clay that can be shaped, we prefer to strive for a more progressive approach in accepting plants as living beings. 

To go beyond a mere automatization of gardening practice, a new plant shaping methodology for the requirements of living architecture needs to be developed and tested based on knowledge and means of plant science. These new methods should also be open for automatization but in addition allow for a bio-hybrid system where plants with their needs and capabilities are accepted in an equal role. In a recent paper, we discuss the biological background for developing robot-plant bio-hybrids~\citep{skrzypczak17}. Tropisms and in particular phototropism (i.e., a plant's growth towards light) are investigated in \textit{flora robotica} for achieving spatial targets by a growing plant~\citep{wahby16a}. Light of different wavelengths influences a plant's growth and development in distinct ways. Hence one stimulator, such as light of RGB LEDs, could direct growth of a tip, provide energy for photosynthesis, guide developmental processes. Moreover, light stimulation exclude a physical contact between a light source and a plant, what reduce touch induce side effects and make easier to design the platform for playing with dynamic, growing structures, such as living plants. We also investigate the application of phytohormones to modify growth direction, architecture, and many physiological and developmental processes in plants (see Fig.~\ref{fig:amu}C). Mechanical stimulation (e.g., vibrations) as in addition to light could modify already grown tissue or the pace of growth (see Fig.~\ref{fig:amu}A and~B). While many of these approaches to manipulate plants may turn out to be specific to certain plant species, some approaches~\citep{wahby16a} could probably have more general applications, especially for closely related plants and for early developmental stages of plants.

Trees have the biggest potential to be valuable components of bio-hybrid living architectural artifacts. Their big size and their ability to keep their shape due to hard tissue in mature stems and branches provide advantageous features. While they are long-lived organisms even after death their wood is useful construction material that can serve as scaffold for young, alive plants. Criteria for selecting the right plant species  are whether they are adapted to the considered environment, their speed of growth, and how susceptible they are to certain stimuli and methods of plant shaping. 

To ease the development of a basic methodology and to simplify experimentation, we currently work not with trees but with climbing plants. They are also a valuable substrate for living architecture. Using climbing plants allows us to use prepared scaffolds to constrain space to a finite number of options for the plant’s directional growth. Some climbing plants also strongly attach to scaffolds and could hence provide a connection mechanism to combine construction materials composed as modules. We believe that monocotyledonous plants (i.e., flowering plants containing only one embryonic leaf) with intercalary meristems (i.e., cells at the stem that allow the plant to grow leaves adaptively at different positions and to regrow after damage) and different types of growth may prove to be very useful. In this case we could be  more efficient  by treating other organs besides the tip meristems to modify the growth direction. The family of monocots contains not only soft grasses and herbs, but also bigger, stiffer plants, such as palm trees (belonging to \textit{Arecaceae}), bamboo (belonging to \textit{Poaceae}), and dracenas ( belonging to \textit{Asparagaceae}). For example, to obtain structures in the conditions of Central Europe, we suggest the use of beech (\textit{Fagus sylvatica}), hornbeam (\textit{Carpinus betulus}), ivy (\textit{Hedera helix}), and \textit{Fallopia augbertii}. However, our preliminary experiments have been conducted mostly with a bean (Phaseolous vulgaris), a poplar (\textit{Populus sp}.), \textit{Arabidopsis thaliana}, \textit{Dracaena marginata}, and a sunflower (\textit{Helianthus annuus}). Currently investigated species were selected for their abilities concerning crucial biological processes, such as long distance signaling.

\section{Architecture}
\begin{figure}[t]
\includegraphics[width=\textwidth]{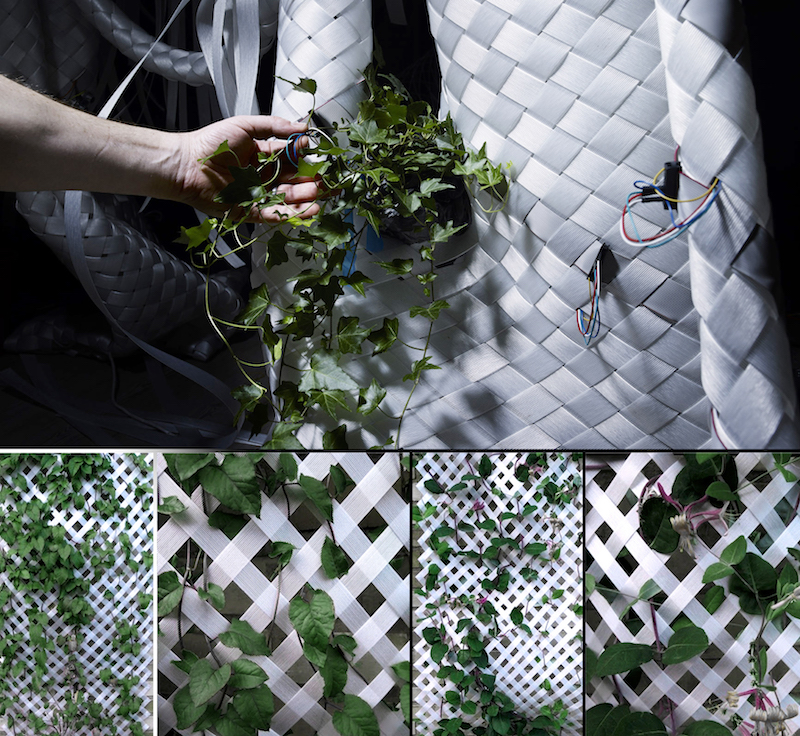}
\caption{\label{fig:cita}Braids as scaffold for plants and sensors.}
\end{figure}
Growing living architecture by combining natural plants with artificial building material and robots requires novel innovative methods in architecture. While robot systems as means to automate construction -- both on site and for pre-fabrication -- have been investigated in depth, using (construction) robots for long periods of time, especially when the building is already occupied, is a novel concept that necessitates new approaches.

Buildings are typically pre-determined, pre-designed, and sometimes even pre-fabricated. Growing living architecture that self-organizes on-site, whether the growth is artificial or biological, requires a categorically distinct approach. Several existing built structures incorporate biological growth on mechanical scaffolds, but use a pre-determined design. These existing approaches first construct an inert scaffold in full, then allow biological growth to fill in the gaps. By contrast, our approach of hybridizing biological growth with a continuous artificial growth process for mechanical scaffolds allows us to move beyond pre-determined design, bringing on-site self-organization into play. We select braid as the overall organizational structure for mechanical scaffolds that are able to grow continuously, due to its many features of flexibility. Because braided structures are secured through their interlacing geometry rather than adhesion or fastenings, they can expand and contract without sacrificing structural performance and can therefore be robotically actuated to adjust their morphology. The filaments of a braided structure can also be unbraided and rebraided, allowing its topology to be continuously re-organized throughout the growth process. Furthermore, braid provides flexibility across scales, materials, and fabrication processes. Braids can then be used to hold immobile robots and to serve as scaffold for climbing plants (see Fig.~\ref{fig:cita}). Other species, such as palm trees, bamboo, and or dracenas can grow within tubular braids to steer and bend them while making sure enough light is provided. The flexibility of the braid concept helps here as the braid scaffolds and tubes can be braided either very stiff to provide proper support or rather loosely to allow for enough incident light.

\section{Braiding robots}

We use braids as (temporary) architectural artifacts in the form of scaffolds, surfaces, tubes, and tree-like shapes to initiate the bio-hybrid system before plant growth of any considerable amount has happened.
To create a braided structure, multiple continuous filaments are interlaced through a structured pattern. We have developed a modular robotic system that can interlace filaments through various patterns, resulting in various topologies of braids. This robot can be used to pre-fabricate braids or to braid, for example, scaffolds at the site possibly directly adapting size and branching of tubular shapes to the currently growing plant. These growth patterns can be influenced by input from sensors and robotic nodes mounted on the scaffold in a generative process, for example, by the Vascular Morphogenesis Controller (VMC).

\begin{figure}[t]
\centering
\includegraphics[width=1.0\textwidth]{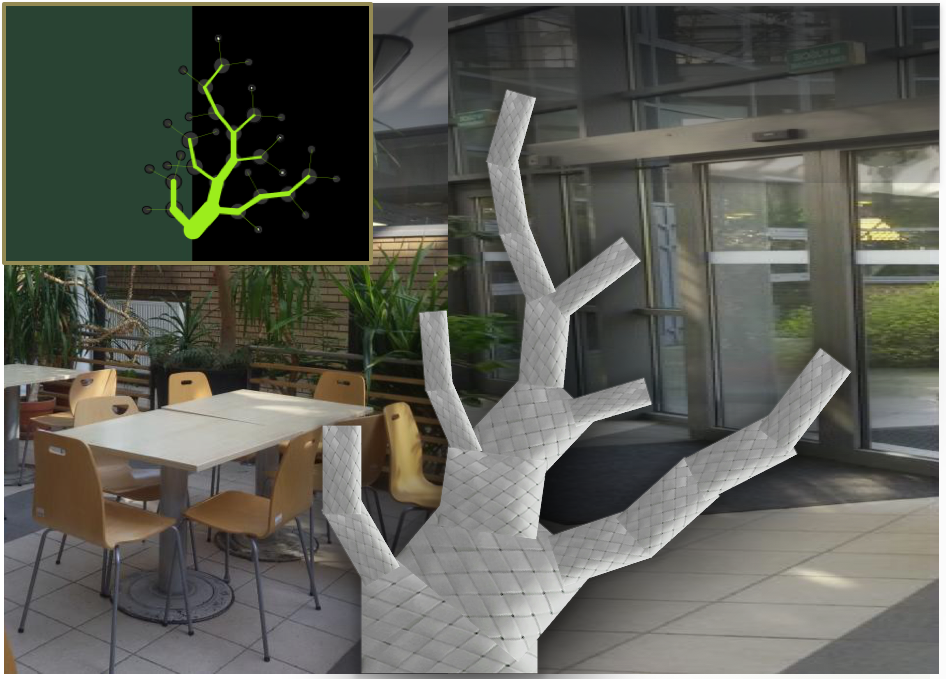}
\caption{\label{fig:vmccombined} An example VMC structure in simulation (inset image) to guide the branching of braids and the number of filaments in each branch. This sketch shows a braided scaffold that is attracted and growing towards the space at the right side because it is infrequently occupied by furniture or people.}
\end{figure}

The VMC~\citep{zahadat17} algorithm was developed based on inspirations from decentralized decision-making mechanisms of branch development in plants. It combines the predetermined rules of branching with global and local sensory information from the environment and leads to a self-organized process of growth for physical structures. 
When scaffolds are mounted at the site and equipped with robotic nodes and sensors, the VMC algorithm running on the robotic nodes can suggest to the braiding robot where and when the branching/fusion of the scaffold should happen and how many of the braiding filaments should go to each branch depending on the current status of the scaffold and the environment (including the plants and humans). Fig.~\ref{fig:vmccombined} illustrates an example braided scaffold where the development of its structure is guided by VMC.

\begin{figure}[t]
\begin{tabular}{cccc}
\subfloat[][Two types of modules make up the braiding machine, driver module (left) and switch module (right).]{\includegraphics[width=0.6\textwidth]{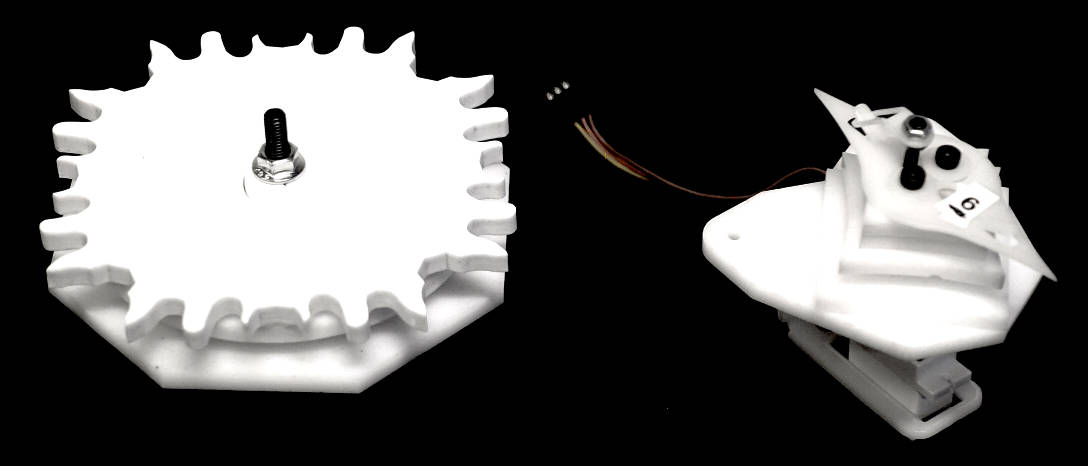}} & \\
\subfloat[][16 Modules assembled as two circles of each 8 modules. The circles are connected over two driver modules, and share one switch module.]{\includegraphics[width=0.6\textwidth]{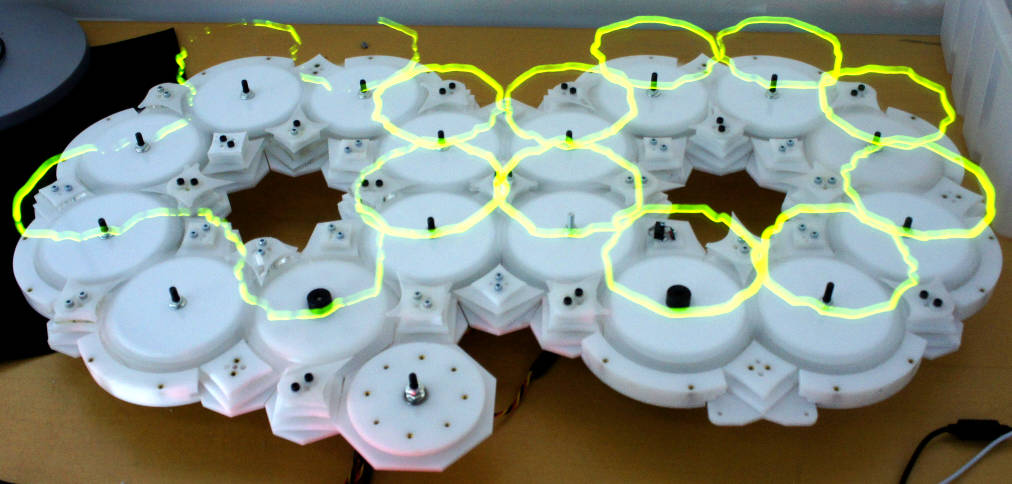}} &
\multirow{-2}[85]{*}{\subfloat[][Filament is connected to dispensers which are driven by the modules to create braided structures.]{\includegraphics[height=0.397\textheight]{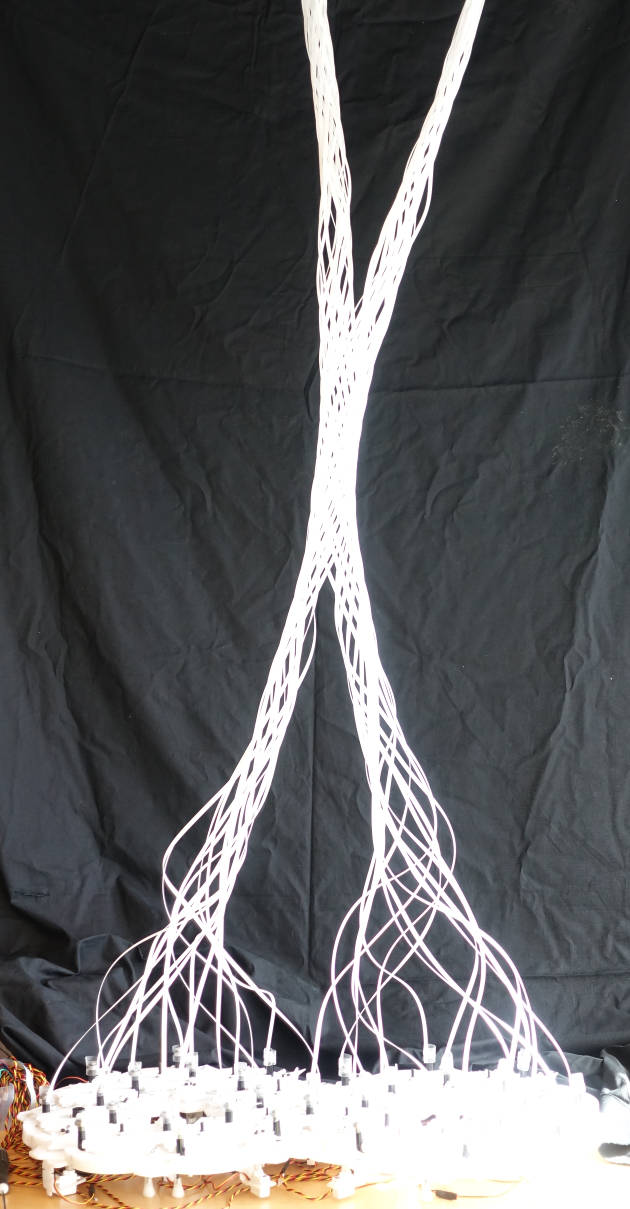}}}
\end{tabular}
\vspace{4mm}
\caption{\label{fig:itu}The modular braiding robot design.}
\end{figure}

The braiding robot system is composed of a software control side and a modular hardware side connected over an integrated hardware-software circuit with sensing and timing control. The hardware consists of two main types of modules: driver modules and switch modules as seen in Fig.~\ref{fig:itu}a. These are laid out on a flat surface and by assembling them one can create strings, circles, and matrices that together compose the braiding machine (see Fig.~\ref{fig:itu}b). In addition to the two module types, other inactive support modules have been designed. The modules can be assembled in different configurations corresponding to the desired braid patterns and topologies. When assembled the modules create a self-intersecting, concave rail system that allows to systematically carry material dispensers by the modules in cooperation. The braided structures are created when many such material dispensers simultaneously extrude filaments while moving in intersecting pathways (see Fig.~\ref{fig:itu}c).

A software side of the system both calculates and controls the collaboration between modules, and ensures that carriers do not collide. This software provides representations that help the operator of the machine both to verify the collaboration and to control loading and unloading of dispensers.

Braided structures are already found in many industrial applications where they prove resilient both in regards to strength and dynamic properties. The advantage of this system over existing industrial systems is the variety of braided topologies, and the capacity to transition from one topology to the other in the same braid structure (see Fig.~\ref{fig:itu}c). That means this modular robotic system is capable of producing a variety of braid morphologies in different materials. Morphologies with different materiality and various dynamic properties may open the way to many additional industrial applications and we imagine these lightweight dynamic structures embedded with sensing, actuating, and computing units as self-contained robotic entities. This way the concept of multiple interacting robots can be extended to a heavily distributed sensing and actuation system that combines concepts from ambient computing and soft robotics~\citep{majidi13,hawkes17}.

\section{Robots and sensors on braids and plants}
In \textit{flora robotica} we want to develop a self-contained bio-hybrid system that keeps robots, sensors, and plants also spatially closely together. This way we have a higher potential to fully integrate the technological components into the grown structures, the sensors actually require vicinity in most cases, and we may even hide the robots behind plants. Similarly, the interactions between robots, sensors, and plants are closer and assigning them equal roles may be easier to achieve. Hence, the robots and sensors are put initially on the braided structures until the plants grow strong enough to hold and carry them.

\begin{figure}[t]
\subfloat[][Robotic node and growing bean.]{\includegraphics[width=0.5\textwidth]{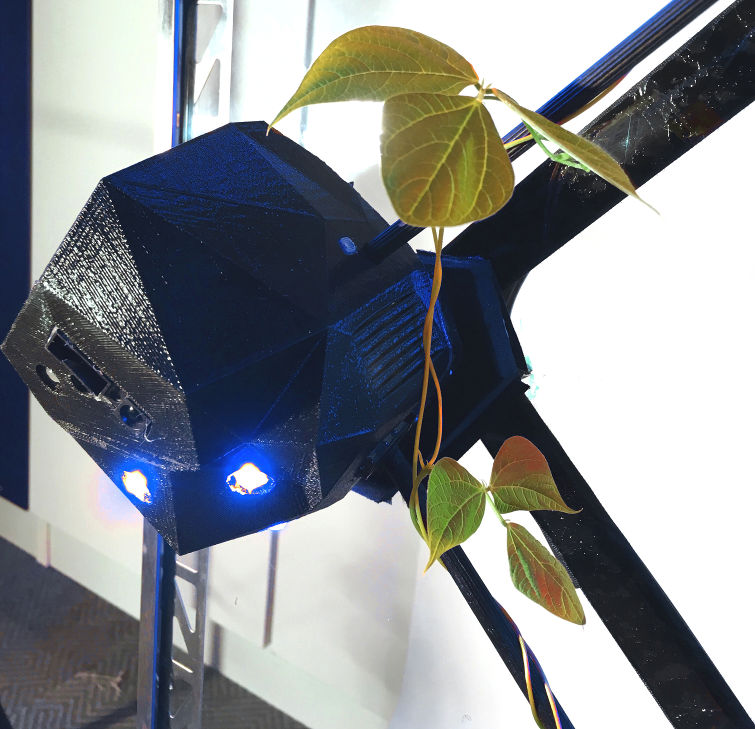}} 
\subfloat[][Sensors and actuators of the robotic node.]{\includegraphics[width=0.5\textwidth]{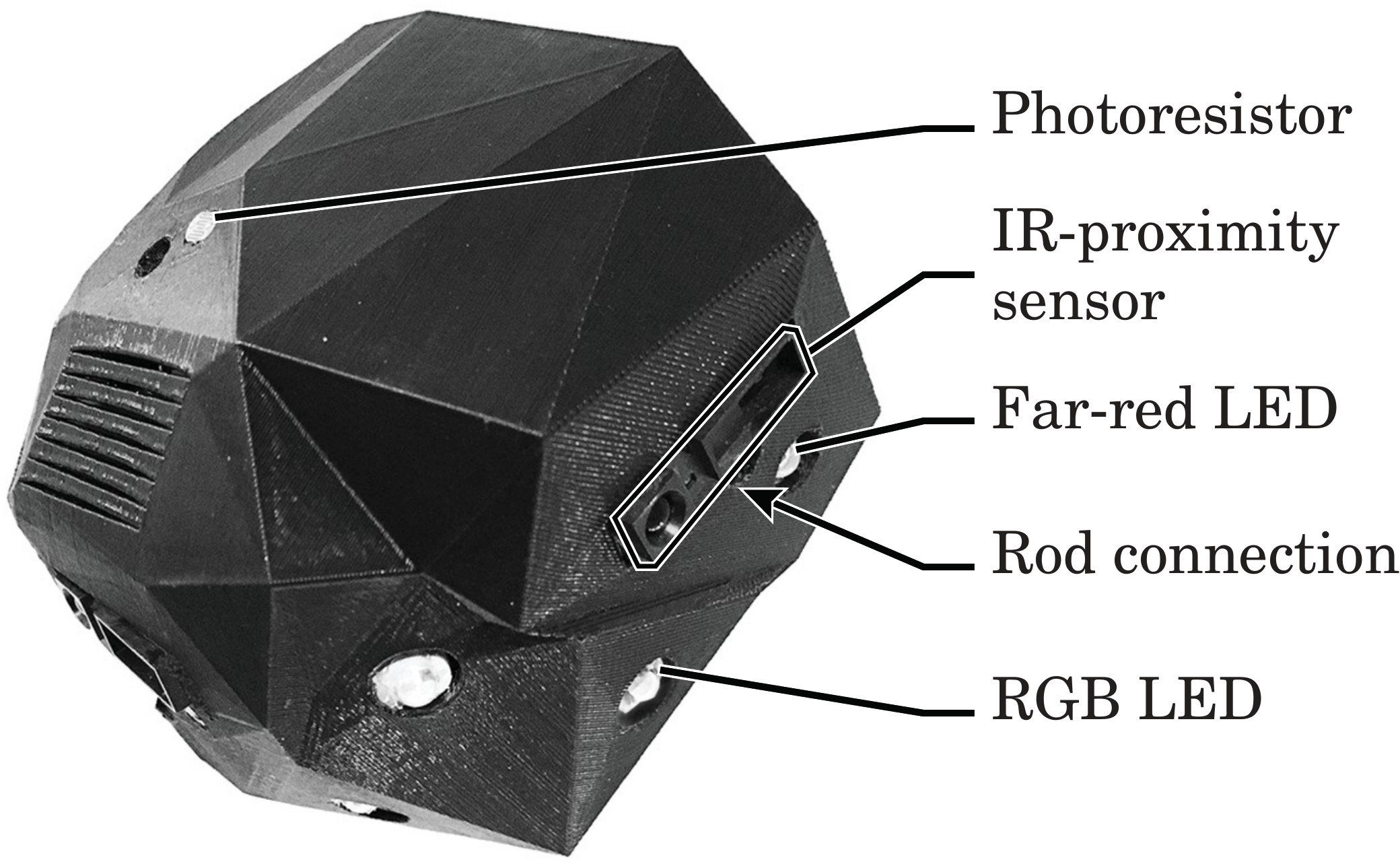}} \\
\subfloat[][Concepts of robotic nodes in braided structures.]{\includegraphics[width=1.0\textwidth]{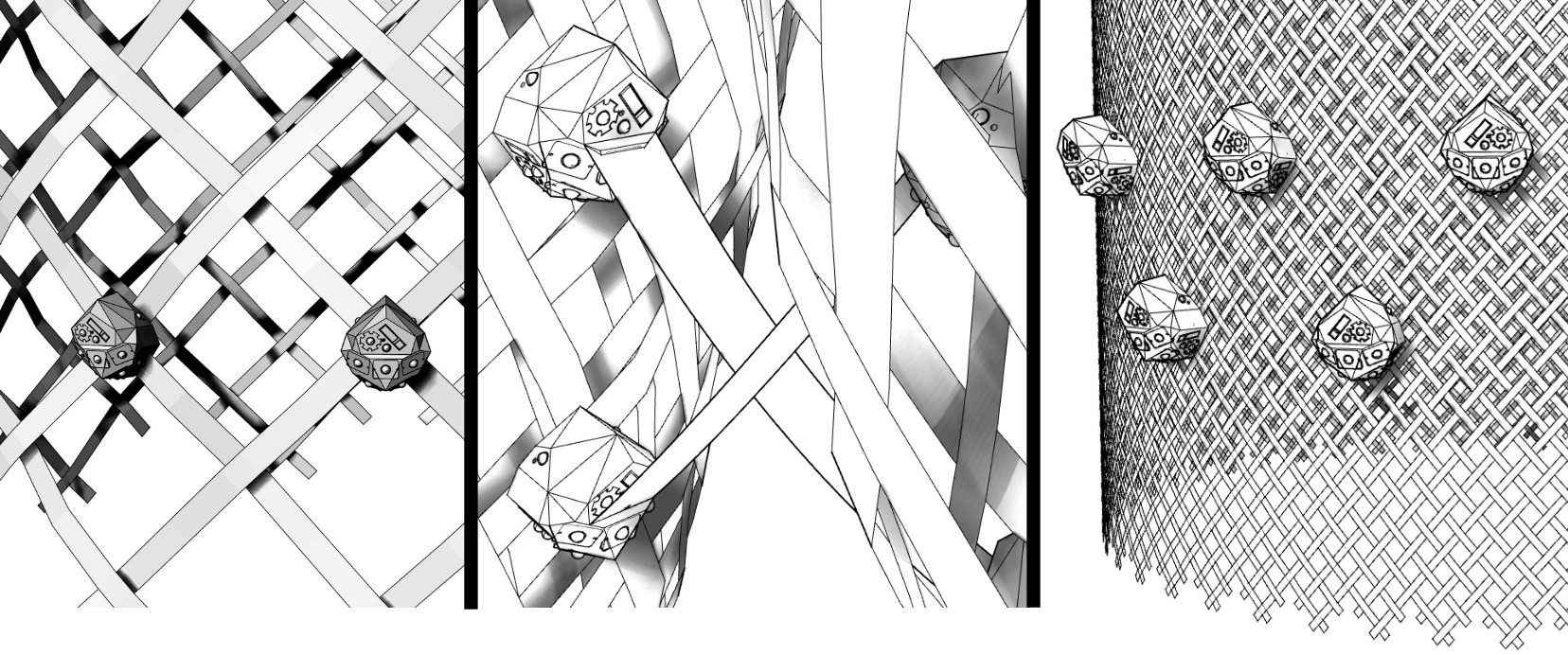}}
\caption{\label{fig:kylo}The robotic node, its components, and visualization of of the concept of robotic nodes in braided scaffolds.}
\end{figure}

In our development of robotic nodes (see Fig.~\ref{fig:kylo}a and~b) that steer directional plant growth, we require integrated concepts of sensing (detecting nearby plants and plant organs) and actuation (in the form of providing appropriate stimuli to the plants). Both requirements are challenging. Plant organs, such as the shoot's growth tip, can have geometries and surfaces that are unfortunate for sensors (small, round, light absorbing). The executed stimuli need to be of correct intensity, as too little triggers no reaction in the plant and too much can stress or even kill the plant. In addition, the robots need to fit into our overall motivation in terms of architectural function and need to be compatible with the braiding approach.

Our distributed robotic nodes are incorporated within the braids (see Fig.~\ref{fig:kylo}c). They steer and shape growth patterns of shoots as the plants mature along mechanical scaffolds. These plant-shaping robots interact with the self-organizing behaviors of plants by robotically providing stimuli that trigger directional growth and motion responses. 

LED actuators in the robots can either trigger a phototropic response and attract the plant, by providing blue light, or can trigger the plant's shade-avoidance syndrome, by providing far-red light. 

Though plants can use the full visual spectrum of light as photosynthetic food source, their irreversible growth towards favorable light conditions -- governed by the phototropism behavior -- is triggered specifically by blue light and UV light. Therefore, while the plants are kept healthy and fed by ambient red light in our setup, a concentrated source of blue light is able to reliably attract the plant and steer the direction of growth. The plants' shade-avoidance behavior is more complex than the relatively straightforward phototropic response. The response is triggered by high concentrations of far-red light (an indicator in natural environments of neighboring plants in close proximity), but can also be impacted by mechanical stimulation and emitted chemicals. When the response occurs, it may cause a combination of behaviors, including both repulsion of growth direction and faster stem elongation. IR-proximity sensors detect the presence of an approaching plant growth tip and photoresistors detect the actuation state of neighboring robotic nodes. 

Using a weighted arithmetic mean approach, we are able to use the IR-proximity sensor to reliably detect approaching plant growth tips as far away as 5~cm. The sensor does not interfere with the shade-avoidance behavior actuated through the far-red LEDs, as no critical wavelength overlap occurs at distances greater than 2~mm from the sensor.
We use a distributed approach to control the robots, so that their artificial self-organization might hybridize with the natural self-organization of the plants. 

Experiments conducted so far demonstrate the ability of the nodes to shape climbing bean plants, steering the plants' binary decisions about growth directions as they navigate a mechanical scaffold.

To create a truly bio-hybrid system, we need to provide two-sided communication between plants and robots, robots and plants, plants and humans. Interfacing with plants is challenging because of their complex biochemistry, their slow responses, and slow  reactions to responses. While the well-being of the plant, its physiological parameters, and its growth mostly depend on environmental conditions, that are easily monitored, we need to gather and provide information about the  plant’s direct responses and reactions to certain stimuli and environmental conditions. Plants are well-known to have a sophisticated and fine ability to sense and respond to changes of environmental parameters and external perturbations. 

Typically, there are several approaches used to measure the plant responses, such as: electrophysiological measurements and temperature/humidity measurements. In the latter case,  we distinguish between transpiration sensing and sap flow sensing.

Electrophysiology is more reliable and more easily applied in typical setups. The electrodes are implemented in the form of silver needles (Ag 925 0.2 mm diameter) that are injected into plant tissue. They are used to provide the electro-stimulation and electro-measuring interface between plants and low-level electronics. We usually select different points at the plant stem to inject the electrodes. There are  two types of measurements: bio-potential measurements and impedance tissue measurements.

\begin{figure}[t]
\centering
\includegraphics[width=0.45\textwidth]{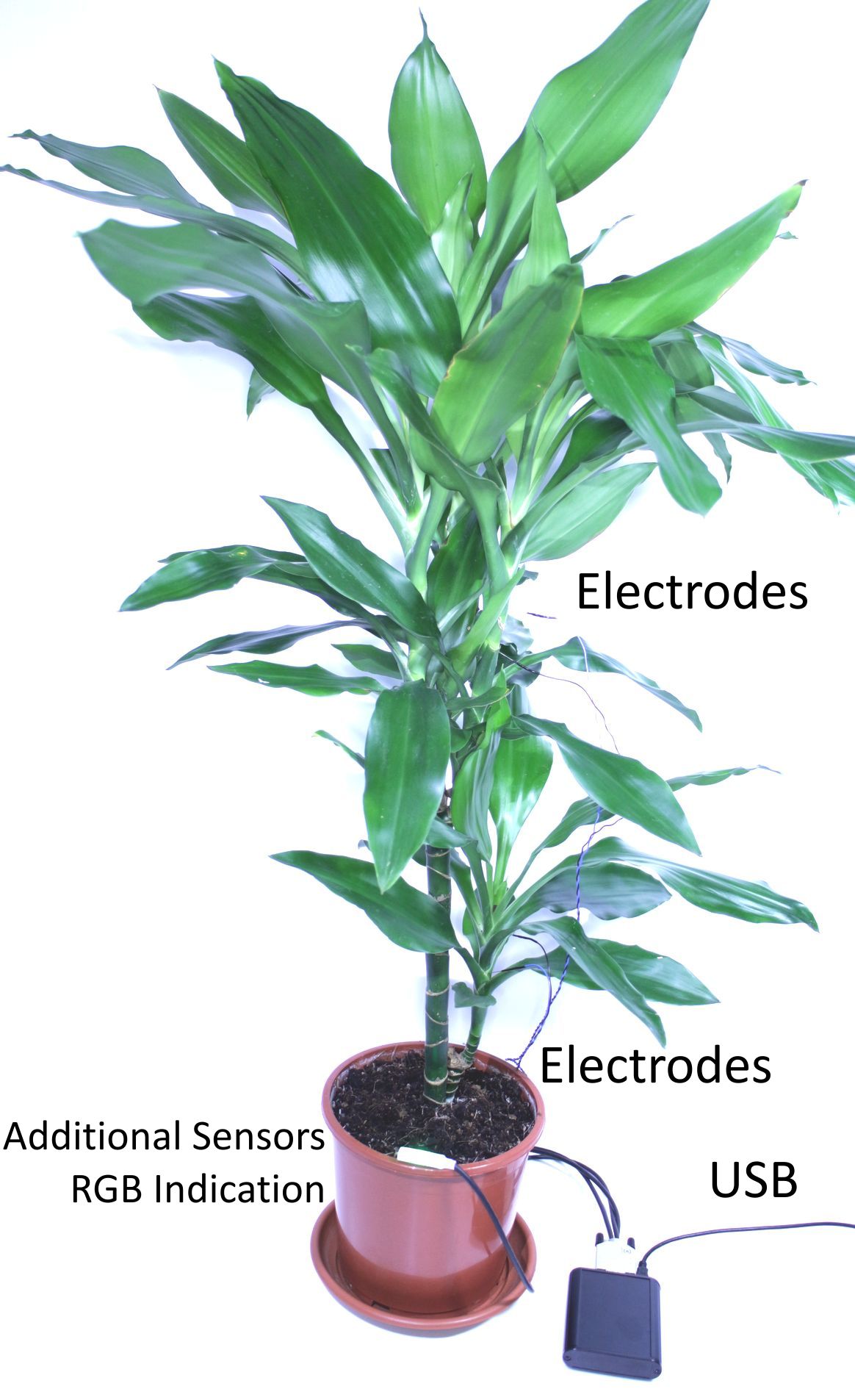}
\caption{\label{fig:cyb}The plant electrophysiology measurement setup.}
\end{figure}

We have developed already several generations of phytosensing systems. The current version, called MU3.3, is shown in Fig.~\ref{fig:cyb}. It includes two channels for electrophysiological measurements, two separate channels for impedance measurements, and two separate channels for bio-potential measurements. The system also includes a transpiration sensor, a sensor that is mounted at the back side of a leaf of a plant to measure the evaporated water. In addition, the phytosensing system includes other sensors for measuring temperature, humidity, ambient light, a 3d accelerometer, a 3d magnetometer, and three channels for actuators.

\section{Putting it all together}

The ultimate goal of \textit{flora robotica} is the creation and exploration of a plant-robot-human ecology in the form of an architectural ensemble. This can be a single entity in the form of an inhibited living architecture or it can be a collection of spatially distributed \textit{flora robotica} systems -- a technologically enhanced social garden. In our project, this so-called Social Garden will be explored in physical and digital instances. The physical presence will incorporate all sites of physical \textit{flora robotica} and connect these over the Internet. 
The digital component of the Social Garden is Internet-based, allows for virtual user interaction, and acts as a searchable data repository of growth patterns, control algorithms, and data tested and acquired at the physical sites.

\subsection{Self-repairing living architecture}

\begin{figure}[t]
\centering
\includegraphics[width=1.0\textwidth]{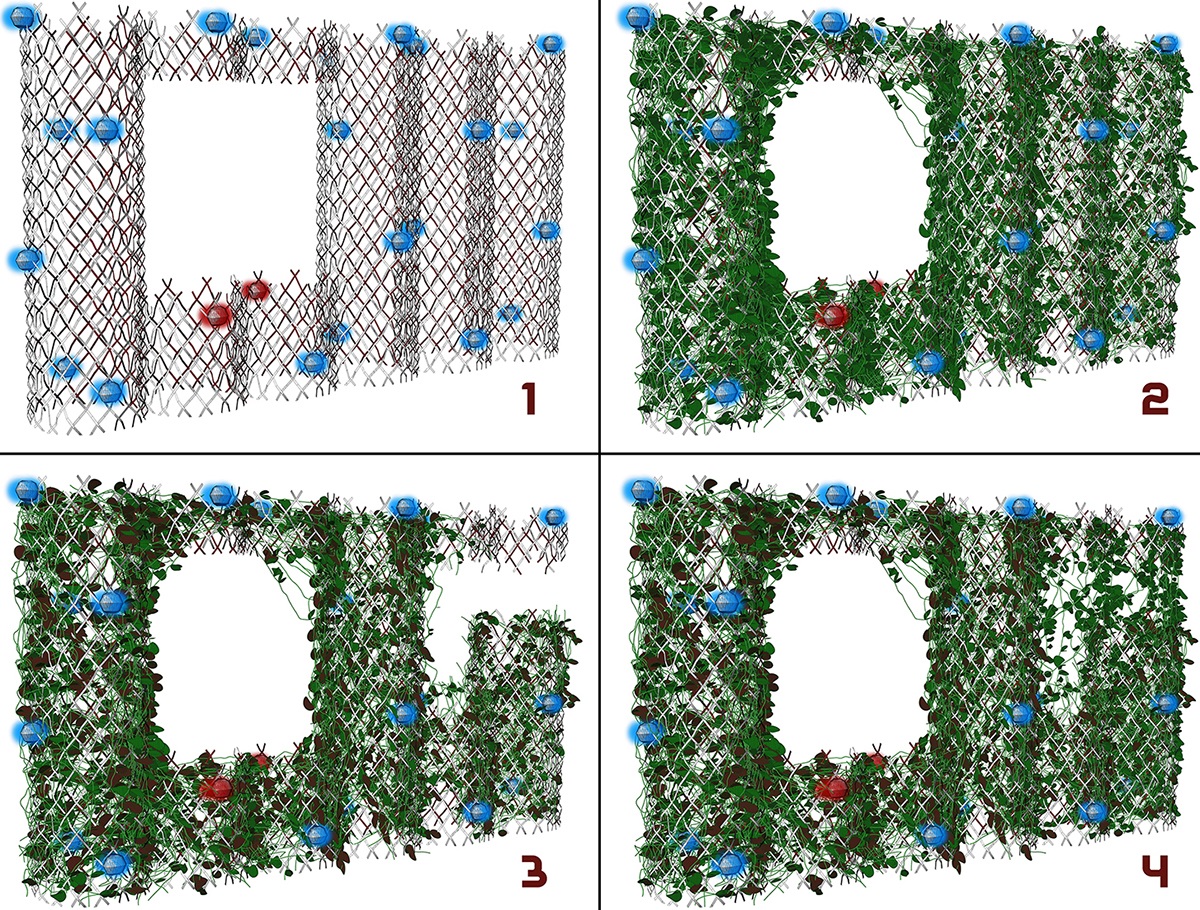}
\caption{\label{fig:benchmark}Sketch of a possible and feasible showcase of \textit{flora robotica}. We combine (1) braided scaffolds, (2) steered plant growth with growth-free opening, (3) removal of material as a test, and (4) subsequent self-repair.}
\end{figure}

Within the project we want to showcase an integrated demonstrator of our approaches on a small scale (see Fig.~\ref{fig:benchmark}). Initially we produce a braided scaffold with the braiding robot supported by the VMC approach and position robotic nodes (Fig.~\ref{fig:benchmark}-1). Appropriately selected plant species are grown along these structures and their growth is steered by the robotic nodes (Fig.~\ref{fig:benchmark}-2). This could include the detection of passages frequently used by humans or the definition of windows that, hence, need to be avoided by the plants. The plants are monitored by our sensors on the braid and the plants. As a benchmark we plan a windowed wall and punch a hole into our grown living architecture (Fig.~\ref{fig:benchmark}-3). The hole needs to be repaired autonomously by the system. Within the window area, plant growth is still prohibited. This benchmark is completed successfully once the hole is self-repaired while the window is still free of plants (Fig.~\ref{fig:benchmark}-4). In this way we would prove one main advantage of living architecture, which is adaptive self-repair on site.


\subsection{Social Garden -- interactive living architecture}

\begin{figure}[t]
\centering
\includegraphics[width=1.0\textwidth]{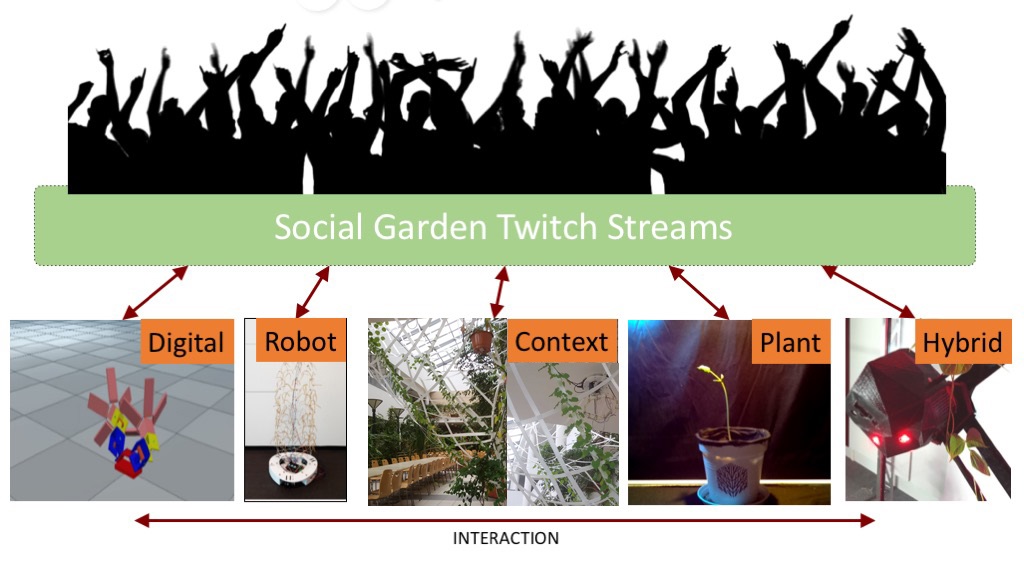}
\vspace*{-5mm}\caption{\label{fig:twitch}Plant-robot interactions augmented by users, who interact with simulations, robots, plants, and architectural artifacts.}
\end{figure}

The idea of the digital presence of \textit{flora robotica} is to connect all our approaches (simulation and plant/robot experiments) and to make our developments available to a wider community. Users can experience the Social Garden remotely, they are attracted to the idea of living architecture, they are allowed to interact with the robots directly, and may share growth patterns and plant growing recipes (see 
Fig.~\ref{fig:twitch}). As key technology to implement an interaction interface is Twitch\footnote{\url{http://twitch.tv}}.

All communication in Twitch is done in the chat interface and this interface can be directly extracted using a chat bot. Our chat-bot is an implementation of a python-based interface originally developed by \citet{anetsberger16}. We can directly extract any message that has been posted in the twitch channels along with the associated user name. In an initial implementation a Twitch user controlled a modular robot with a varying amount of modules. The user was also allowed to request changes of the robot's morphology. In ongoing research, we work on approaches that allow users to interact with braided robots and even plants. 

\section{Discussion and Conclusion}

In our effort towards a generic basic methodology to grow plant-robot living architecture we have developed a complete, consistent approach and we have found effective experimental setups to use sensors and immobile robots to steer the directional growth of natural plants. We have selected appropriate plant species and we continue to experiment with additional species and stimuli (e.g., hormones, vibrations). We develop novel architectural approaches to address the unconventional setting of \textit{flora robotica}. Braiding as our leading technique has opened up a vast variety of opportunities for designs and to combine plants with robot-produced structures. We continue our development of the braiding robot integrated with VMC, that provides the desired shapes, and other techniques of predicting the resulting morphologies and their properties. The design of the robotic nodes, that we have developed, will be refined and we will continue with experiments to steer plant growth in more and more complex shapes.

Our approach of combining plants with technical architectural elements is conceptually similar to prior work~\citep{ludwig2016designing} but extends this in significant ways. First, we try to establish a symbiotic coupling of plants and robots. Second, we produce scaffolds using braid technique that allows for  modification, either manually or self-organized through robotic fabrication. Third, we develop and implement robotic nodes and algorithms that control the adaptation and growth of the bio-hybrid in response to its long-term environment and resource balancing.

In the second half of the project, we plan to refine our approaches and to showcase them in a demonstrator. Large-scale tests and diversification of the techniques will be left for the farther future. The concept of living architecture holds potential for influencing how we construct our built environment in future with a greater consideration for material accumulation, resource balancing, environmental respect and adaptation to changes in spatial needs. Growing parts of our future houses, or even complete houses, with natural plants could help make our cities healthier, greener, and more livable.

\section*{Acknowledgment}
Project `{\it flora robotica}' has received funding from the European Union's
Horizon 2020 research and innovation program under the FET grant
agreement, no.~640959.

\bibliographystyle{plainnat}
\bibliography{bib}

\end{document}